\documentstyle[aas2pp4]{article}
\newcommand{\bc}{\begin{center}}
\newcommand{\ec}{\end{center}}

\def\ltsima{$\; \buildrel < \over \sim \;$}
\def\ltsim{\lower.5ex\hbox{\ltsima}}
\def\gtsima{$\; \buildrel > \over \sim \;$}
\def\gtsim{\lower.5ex\hbox{\gtsima}}

\def\frac#1#2{{#1\over#2}}           
\begin{document}

\title{THE ABSOLUTE MAGNITUDE OF RR LYRAE STARS DERIVED FROM 
THE HIPPARCOS CATALOGUE}

\author{Takuji Tsujimoto$^{1}$, Masanori Miyamoto$^{1}$, Yuzuru
Yoshii$^{2,3}$}

\vspace{0.5cm}
\affil{$^1$National Astronomical Observatory,
         Mitaka, Tokyo 181, Japan}
\affil{$^2$Institute of Astronomy, Faculty of Science,
         University of Tokyo, Mitaka, Tokyo 181, Japan}
\affil{$^3$Research Center for the Early Universe, School
         of Science,
         University of Tokyo, Bunkyo-ku, Tokyo 113, Japan}

\begin{abstract}

The present determination of the absolute magnitude $M_V({\rm RR})$ of
RR Lyrae stars is twofold, relying upon Hipparcos proper motions and
trigonometric parallaxes separately.  First, applying the statistical
parallax method to the proper motions, we find $<M_V({\rm
RR})>=0.69\pm0.10$ for 99 halo RR Lyraes with $<$[Fe/H]$>$ =--1.58.
Second, applying the Lutz-Kelker correction to the RR Lyrae HIP95497
with the most accurately measured parallax, we obtain $M_V({\rm
RR})$=(0.58--0.68)$^{+0.28}_{-0.31}$ at [Fe/H]=--1.6. Furthermore,
allowing full use of low accuracy and negative parallaxes as well for
125 RR Lyraes with -- 2.49$\leq$[Fe/H]$\leq$0.07, the maximum
likelihood estimation yields the relation, $M_V({\rm
RR})$=(0.59$\pm$0.37)+(0.20$\pm$0.63)([Fe/H]+1.60), which formally
agrees with the recent preferred relation. The same estimation yields
again $<M_V({\rm RR})>$ = $0.65\pm0.33$ for the 99 halo RR
Lyraes. Although the formal errors in the latter three parallax
estimates are rather large, all of the four results suggest the
fainter absolute magnitude, $M_V({\rm RR})$$\approx$0.6--0.7 at
[Fe/H]=--1.6.  The present results still provide the lower limit on
the age of the universe which is inconsistent with a flat,
matter-dominated universe and current estimates of the Hubble
constant.

\end{abstract}


\section{INTRODUCTION}

The absolute magnitude of RR Lyrae stars $M_V({\rm RR})$ is an
indicator of distance and age in the Galaxy.  In particular it is a
key parameter for determining the age of Galactic globular clusters
(GCs) which should place a lower limit on the age of the universe.
However, the values of $M_V({\rm RR})$ reported so far defies a
consensus and splits into faint and bright values.

The faint value of $M_V({\rm RR})$=0.70--0.75 at the characteristic
halo metallicity [Fe/H]=--1.6, which gives a shorter distance scale or
larger age of clusters, is derived from halo RR Lyraes in the fields
by the statistical parallax method (Hawley et al.~1986; Barnes \&
Hawley 1986; Strugnell, Reid, \& Murray 1986; Layden et al.~1996) and
by the Baade-Wesselink analysis (Liu \& Janes 1990a; Jones et
al.~1992).  On the other hand, the bright value of $M_V({\rm
RR})$=0.44 at [Fe/H]=--1.9, which gives a longer distance scale or
smaller age, is derived from RR Lyraes in the GCs of the LMC, assuming
a distance modulus to the LMC to be 18.5 (Walker 1992).  Sandage's
pulsation theory (Sandage 1993) and RR Lyraes in the metal-poor GCs of
M5 and M92 (Storm, Carney, \& Latham 1994) also support the bright
value, but this is not the case for RR Lyares in M4 (Liu \& Janes
1990b).

Such a dichotomy in current estimates of $M_V$({\rm RR}) is the
dominant source of uncertainty in the age determination of GCs.  The
faint $M_V({\rm RR})$ raises a well-known problem such that the age of
GCs exceeds the age of the universe if the Hubble constant is adopted
as $H_0=60-80$km$\,$s$^{-1}\,$Mpc$^{-1}$, and this is the central
argument in favor of nonzero cosmological constant (Bolte \& Hogan
1995).  However, a possibility of swaying back to the bright $M_V({\rm
RR})$ has recently been discussed since the preference of its bright
value was inferred from the Hipparcos measurements of parallaxes of
Cepheids (Feast \& Catchpole 1997) and subdwarfs (Reid 1997).

The astrometric observations of field RR Lyraes have been available to
us as a part of the Hipparcos programme assigned to the proposers in
1982.  This paper is therefore the first report in the main journal on
the {\it direct} determination of $M_V({\rm RR})$ from statistical
treatments of Hipparcos proper motions and trigonometric parallaxes of
RR Lyraes (cf., Tsujimoto, Miyamoto, \& Yoshii 1997; Fernley et
al.~1997).  The derived results from the proper motions in \S 2 and
from the parallaxes in \S 3 have confirmed the faint $M_V({\rm RR})$,
thus suggesting that the so-called age discrepancy would not be
resolved or alleviated yet.

\section{$M_V({\rm RR})$ DERIVED FROM HIPPARCOS PROPER MOTIONS}

The Hipparcos proper motion system is inertial, on a global sense,
within the error $\pm$ 0.25 mas/yr and the mean error
$(\sigma_\mu)_{\rm H}$ of the Hipparcos proper motions
[$(\mu_\alpha^*)_{\rm H}$, $(\mu_\delta)_{\rm H}$] for the RR Lyraes
considered here is [$\pm$3 mas/yr, $\pm$2 mas/yr]. Thus, the previous
proper motions are remarkably improved by the Hipparcos mission. The
most essential part of improvements of the previous proper motions
[$(\mu_\alpha^*)_{\rm p}$, $(\mu_\delta)_{\rm p}$] may be described by
the rotation vector \mbox{\boldmath$\Omega$} of the previous proper
motion system with respect to the Hipparcos system. The vector
components ($\Omega_x$, $\Omega_y$, $\Omega_z$) in the equatorial
rectangular coordinates are derived from the equations of condition:

\begin{math}
\begin{array}{lclcl}
[(\mu_\alpha^*)_{\rm p} - (\mu_\alpha^*)_{\rm H}]&=&-\Omega_x \sin\delta 
\cos\alpha & - & \Omega_y \sin\delta\sin\alpha \\

[(\mu_\delta)_{\rm p} - (\mu_\delta)_{\rm H}]&=&+\Omega_x \sin\alpha
& - & \Omega_y \cos\alpha \\
\end{array}
\end{math}

\begin{equation}
\left.
\begin{array}{l}
\hspace{2.cm}+\Omega_z \cos\delta \\
 \\ 
\end{array}
\right\} \ \ \ .
\end{equation}

Applying eq.[1] to the proper motions of 99 halo RR Lyraes with
$<$[Fe/H]$>$= --1.58, which are common to Table 2 compiled by Layden
et al.~(1996) and Hipparcos Catalogue, we have

\begin{equation}
\left.
\begin{array}{rcr}
\Omega_x & = & +1.61\pm0.15 {\rm mas/yr} \\
\Omega_y & = & +4.31\pm0.13 {\rm mas/yr} \\
\Omega_z & = & -1.27\pm0.13 {\rm mas/yr} \\
\end{array}
\right\} \ \ \ .
\end{equation}

Thus, the previous proper motions are systematically improved by the
amount of $|\mbox{\boldmath$\Omega$}| \sim 5$ mas/yr larger than the
mean error $(\sigma_\mu)_{\rm H}$ in Hippracos proper motions for the
RR Lyraes. Encouraged by the clear improvement indicated by eq.[2],
we apply the statistical parallax method to Hipparcos proper motions
to improve $M_V({\rm RR})$ previously determined.

\begin{table*}
  \caption{\em Results of statistical parallax analysis
}
  \label{tab:table}
  \begin{center}
    \leavevmode
    \footnotesize
    \begin{tabular}[h]{cccccccccc}
      \hline \\[-5pt]
           & $N_{\rm stars}$ & $<$[Fe/H]$>$ & $U_\odot$ & $V_\odot$ 
           & $W_\odot$ & $\sigma_U$ 
           & $\sigma_V$ & $\sigma_W$ & $M_V({\rm RR})$ \\[+5pt]
      \hline \\[-5pt]
      Hawley et al.~(1986) & 77  & -- & +21$\pm$19 & --184$\pm$17 
      & -4$\pm$11 
      & 166$\pm$76 & 114$\pm$52 & 91$\pm$40 & 0.73$\pm$0.18 \\
      Layden et al.~(1996) & 162  & --1.61 & --9$\pm$14 & --210$\pm$12 
      & -12$\pm$8 
      & 168$\pm$13 & 102$\pm$8 & 97$\pm$7 & 0.71$\pm$0.12 \\
      Fernley et al.~(1997)  & 69  & --1.66 & -- & --  & --
      & -- & -- & -- & 0.73$\pm$0.18 \\
      this work  & 99  & --1.58 & --12$\pm$17 & --200$\pm$11 & +2$\pm$3 
      & 161$\pm$13 & 105$\pm$9 & 88$\pm$7 & 0.69$\pm$0.10 \\
      \hline \\
      \end{tabular}
  \end{center}
\end{table*}

The method is a direct application of well-understood kinematics of
sample stars to the observations (proper motions, radial velocities,
and apparent magnitudes), when a single luminosity class of sample
stars can be specified within a small intrinsic magnitude
dispersion. The present maximum likelihood statistical analysis of RR
Lyraes is similar to the one done on RR Lyraes by Hawley et
al.~(1986).  Introducing a model that predicts the spatial motion of
each star at each position in the Galaxy, we estimate the model
parameters by the maximum likelihood method. The present model
includes a reflex solar motion with respect to the galactic center and
an ellipsoidal velocity distribution for RR Lyraes, but neglects the
differential galactic rotation. Assuming that the velocity residuals
\mbox{\boldmath$\nu$}, the observed velocity minus the velocity
expected from the model, follow a Gaussian distribution with zero
mean, we have the logarithmic likelihood $\ln L$ of the aggregate of
all the residuals:

\begin{equation}
\ln L = -\frac{1}{2}\Sigma \ln (|\mbox{\boldmath$M$}| + \mbox{\boldmath$\nu'$}
\mbox{\boldmath$M$}^{-1}\mbox{\boldmath$\nu$}) + {\rm constant} \ \ ,
\end{equation} 

\noindent where \mbox{\boldmath$M$} is the covariance tensor given by
the expectation
\mbox{\boldmath$M$}=$<$\mbox{\boldmath$\nu$}\mbox{\boldmath$\nu'$}$>$.

The model parameters describe the velocity components
($U_\odot$,$V_\odot$,$W_\odot$) of the reflex solar motion with $U$,
$V$, and $W$ axes pointing to the galactic center, the galactic
rotation, and the north galactic pole, respectively, the velocity
dispersion components ($\sigma_U$,$\sigma_V$,$\sigma_W$), the
covariances ($\sigma_{UV}$,$\sigma_{UW}$,$\sigma_{VW}$), the absolute
magnitude $M_V({\rm RR})$, and the intrinsic magnitude dispersion
$\sigma_M$ in $M_V({\rm RR})$. The observational errors $\sigma_{V_r}$
in the radial velocity and ($\sigma_{\mu_\alpha}$,
$\sigma_{\mu_\delta}$) in the proper motions for each star are taken
from Layden (1996) and Hipparcos Catalogue, respectively. The
intrinsic dispersion $\sigma_M$ is parameterized by $\sigma_k$, the
dispersion in the distance scale parameter $k$. Here we assume
$\sigma_k$=0.1 (Layden et al.~1996). The present model therefore
incorporates 10 parameters, which are solved by the numerical
minimization technique of simplex optimization.

Combining Hipparcos proper motions with the data of [Fe/H], radial
velocity, apparent magnitude, and interstellar extinction taken from
Layden et al.~(1996) and Layden (1994), we have carried out the
present analysis for 99 halo RR Lyraes with $<$[Fe/H]$>$=--1.58, and
found that all of the covariances
($\sigma_{UV}$,$\sigma_{UW}$,$\sigma_{VW}$) are nearly equal to
zero. Therefore, we have repeated the similar analysis to determine 7
parameters, putting all of the covariances equal to zero. The results
are shown in Table 1.  We find $<M_V({\rm RR})>$ = 0.69$\pm$0.10
together with the solar motion with respect to the galactic center
($U_\odot$,$V_\odot$,$W_\odot$) = (--12$\pm$17,--200$\pm$11,+2$\pm$3)
km s$^{-1}$ and the velocity dispersions
($\sigma_U$,$\sigma_V$,$\sigma_W$) = (161$\pm$13,105$\pm$9,88$\pm$7)
km s$^{-1}$. The RR Lyrae stars as a whole rotate with
$V_{rot}$$\approx$30km s$^{-1}$, assuming the local standard of rest
$V_{\rm LSR}$=220 km s$^{-1}$ and the solar motion of 16.5 km s$^{-1}$
in the direction $l$=53$^\circ$ and $b$=25$^\circ$. These results are
in good agreement with those recently obtained by Hawley et
al.~(1986), Layden et al.~(1996), and Fernley et al.~(1997) (see Table
1) within the error estimates. Thus, the absolute magnitude of RR
Lyraes derived from Hipparcos proper motions, $<M_V>\approx 0.70$ at
$<$[Fe/H]$>$ $\approx$ --1.6, confirms the recent results from the
ground-based observations.

In general, a linear relation between $M_V({\rm RR})$ and [Fe/H] has
been assumed. The slope of $M_V({\rm RR})$--[Fe/H] relation has a
strong influence on the inferred age differences among the globular
clusters, which give a measure of the dynamical timescale for halo
formation. The Baade-Wesselink results give the values of $M_V({\rm
RR})$ over the range of [Fe/H]= --2.2 to 0, which are fitted by the
relation: $M_V({\rm RR})$ = (1.02$\pm$0.03)+(0.16$\pm$0.03)[Fe/H] (Jones
et al.~1992) or $M_V({\rm RR})$ =(1.04$\pm$0.10)+(0.21$\pm$0.05)[Fe/H]
(Skillen et al.~1992).  However, the controversial situation holds
also for determining the slope of $M_V({\rm RR})$--[Fe/H]
relation. Sandage's pulsation theory gives much steeper slope, whereas
statistical parallax analyses (Hawley et al.~1986; Strugnell et
al.~1986; Layden et al.~1996) support very weak (or no) dependence of
$M_V({\rm RR})$ on [Fe/H].  It is, of course, possible to incorporate
the linear relation between $M_V({\rm RR})$ and [Fe/H] into the
present analysis by dividing the available sample into bins of [Fe/H],
but the available number (26) of Hipparcos RR Lyraes with [Fe/H] $>$ --
1.0 is too small to yield the reliable relation over the wide range of
[Fe/H].

\section{$M_V({\rm RR})$ DERIVED FROM HIPPARCOS PARALLAXES} 

\begin{figure}[h]
\special{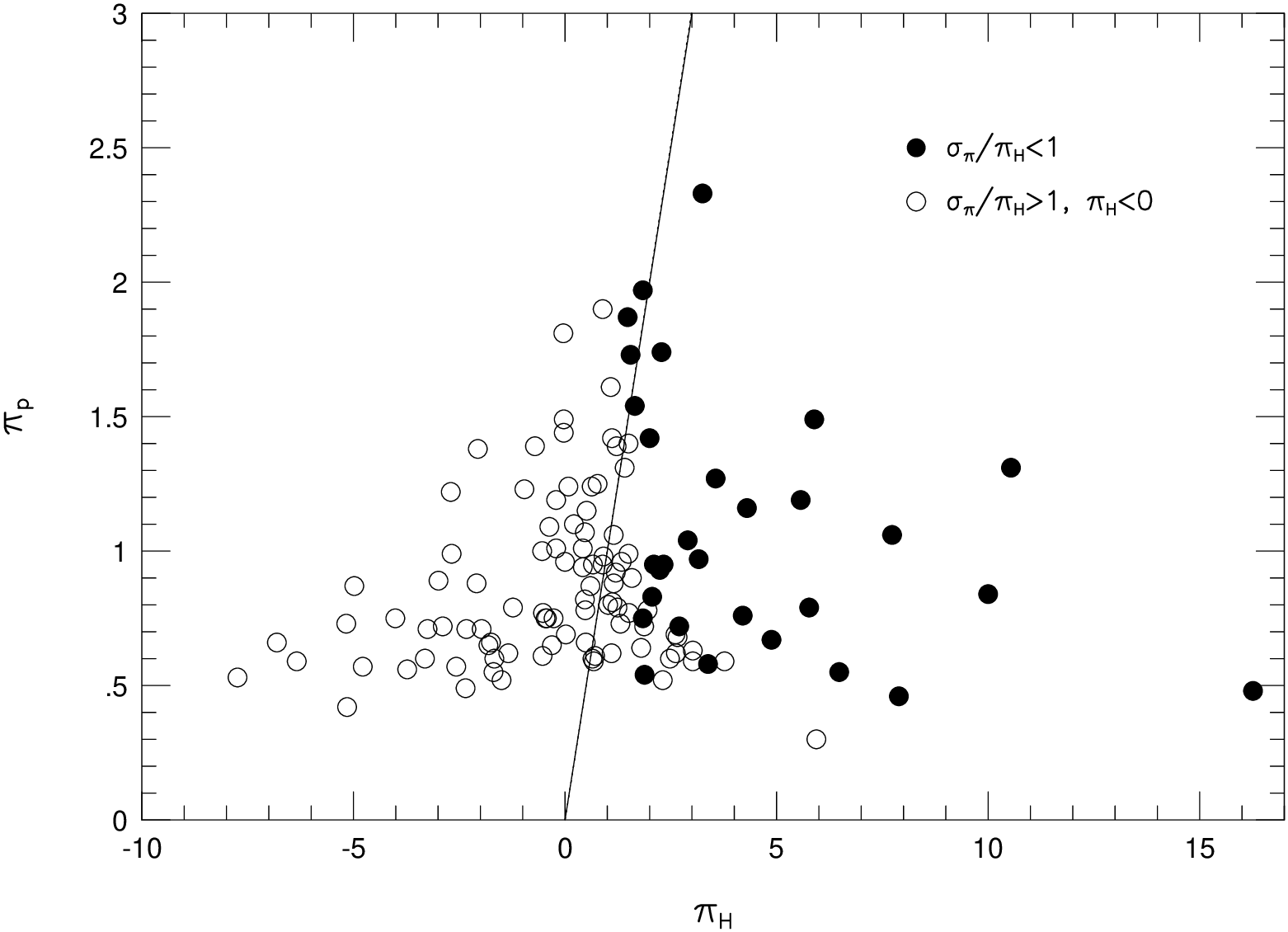 voffset=-250 vscale=40 hscale=35 
hoffset=-20}
\vspace{8cm}
\caption{ The comparison between Hipparcos parallaxes $\pi_{\rm H}$ 
of 125 RR Lyraes and the photometric parallaxes  
$\pi_{\rm p}$ (see eq.[5]). The straight line indicates
the relation $\pi_{\rm H}$=$\pi_{\rm p}$.  }
\end{figure}

The relative error $\sigma_\pi/\pi_{\rm H}$ for 173 Hipparcos RR
Lyraes with the standard error $\sigma_\pi$ of Hipparcos parallaxes
$\pi_{\rm H}$ distributes such that each number of stars with
$\sigma_\pi/\pi_{\rm H}$ less than 1 and larger than 1 is 51 and 65,
respectively. The remaining 57 stars have even negative
parallaxes. Thus, almost all of the parallaxes are measured with very
large error, except one case (HIP95497=RR Lyr) with
$\sigma_\pi/\pi_{\rm H}$ = 0.135. In Figure 1 Hipparcos parallaxes
$\pi_{\rm H}$ are compared with the photometric ones $\pi_{\rm p}$
defined by $r_{\rm p} {\rm (kpc)}=10^{0.2(V-M_V({\rm RR})-10)}$, where
$V$ is the apparent magnitude corrected for interstellar extinction
and $M_V({\rm RR})$ the calibrated absolute magnitude given by the
relation
\begin{equation}
M_V({\rm RR})=a+b([{\rm Fe/H}]-c)
\end{equation}

\noindent with $a$=0.91, $b$=0.20, $c$=--1.60 (see eq.[5]). The filled
and open circles in the figure indicate the parallaxes with
$\sigma_\pi/\pi_{\rm H}$ smaller than 1 and others, respectively. It
is noticed that the individual deviation of Hipparcos parallaxes from
the photometric ones is extremely large, but Hipparcos parallaxes
including low quality and negative ones as a whole agree with the
photometric ones.

The statistical properties of Hipparcos parallaxes of their errors
have been investigated extensively by Arenou et al.~(1995), who have
shown that Hipparcos parallaxes are free from any global zero-point
error up to $\pm 0.1$ mas and their errors are normally distributed
with zero mean. This means that Hipparcos parallaxes including even
the negative ones for any ensemble of stars are statistically
well-behaved. This is the case for the ensemble of RR Lyraes as well,
as is shown in Figure 1.  These statistical regularities of Hipparcos
parallaxes suggest a relevant technique for estimating statistically
the absolute magnitude $M_V({\rm RR})$ of RR Lyraes to be applicable
to the ensemble of RR Lyrae parallaxes, even though each individual
parallax measured is largely erroneous, giving negative one as well.
Here we attempt to derive $M_V({\rm RR})$ directly from Hipparcos
parallaxes, relying upon two statistical methods to find the unbiased
estimate.

a) Lutz-Kelker Correction 

Lutz \& Kelker (1973) have shown that only when the relative error
$\sigma_\pi/\pi'$ is smaller than 0.175, an unbiased estimate of the
absolute magnitude can be retrieved from the measured parallaxes
$\pi'$.  We have only one case of HIP95497 with $\sigma_\pi/\pi_{\rm
H}=0.135$, to which Lutz-Kelker correction is applicable. The
uncorrected absolute magnitude of this star is $M_V$(RR) =
$0.82^{+0.28}_{-0.31}$, which is derived formally from the measured
parallax. On the basis of Hanson's (1979) formulation to estimate the
correction, we have the calibrated absolute magnitude range $M_V({\rm
RR})$ = (0.62--0.72)$^{+0.28}_{-0.31}$ for HIP95497 ([Fe/H]=--1.37),
corresponding to the assumed spatial density of RR Lyraes
$n(r)$=const. and $n(r)\propto r^{-2}$ as the lower and upper bounds,
respectively. If we assume $b$=0.20 in eq.[4], the above range of
$M_V({\rm RR})$ for HIP95497 corresponds to $M_V({\rm RR})$ =
(0.58--0.68)$^{+0.28}_{-0.31}$ at [Fe/H]=--1.6.

b) Maximum Likelihood Calibration

In order to retrieve statistically an unbiased estimate of the
absolute magnitude together with its intrinsic dispersion for a single
luminosity class of magnitude limited stars from their measured
parallaxes with errors, Smith (1987,88) has proposed a
maximum likelihood principle for correcting simultaneously the
Malmquist bias and the Lutz-Kelker bias, allowing full use of low
accuracy and negative parallaxes.

In line with the above principle we attempt to find the best esimate
of the absolute magnitude $M_V({\rm RR})$ and of the intrinsic
magnitude dispersion $\sigma_M$ for 125 Hipparcos RR Lyraes in the
metallicity range --2.49$\leq$[Fe/H]$\leq$0.07. We use here the
numerical algorithm developed by Ratunatunga \& Casertano (1991).
Maximizing the likelihood function given by the product of the
probabilities $p(\pi_{\rm H})$ of all the RR Lyraes considered, we
obtain the best estimate of $\sigma_M$, $a$ and $b$ in eq.[4], where
we adopt $c$ = --1.60, taking into account the fact that the present
distribution of RR Lyraes shows a peak at [Fe/H]$\approx$--1.6.
Setting the allowable range of the measured parallaxes as [$\pi_{\rm
lower}$,$\pi_{\rm upper}$]=[$-\infty$,$+\infty$], and assuming the
number density of RR Lyraes in space to be constant, we find the
$M_V({\rm RR})$-[Fe/H] relation:
\begin{equation}
M_V({\rm RR})=(0.59\pm0.37)+(0.20\pm0.63)([{\rm Fe/H}]+1.60)
\end{equation}
with $\sigma_M=2.6\times10^{-4}\pm0.29$. It is noticed that although
the errors in the parameters estimated are large, the present
$M_V({\rm RR})$-[Fe/H] relation shows a formal agreement with the preferred
relation $M_V({\rm RR})=0.98+0.20[{\rm Fe/H}]$ proposed by Chaboyer,
Demarque, \& Sarajedini (1996) without quoting error estimates. The
insensitivity of the relation given by eq.[5] has been checked by
putting $\sigma_M$ = 0.1--0.2 (Layden et al.~1996), [$\pi_{\rm
lower}$,$\pi_{\rm upper}$]=[0,$+\infty$], and $n(r)\propto r^{-2}$
(Reid 1997), respectively.

The large standard errors in eq.[5] are owing to the small numbers of
degrees of freedom in the present statistics. To increase the freedom,
putting $b$=0 in eq.[4], we apply the same method to the 99 halo RR
Lyraes with $<$[Fe/H]$>$=--1.58, whose absolute magnitude has been
determined already on the basis of the statistical parallax
method. Then, we find $<M_V({\rm RR})>$=0.65$\pm$0.33 and
$\sigma_M$=4.1$\times10^{-6}\pm0.39$ at $<$[Fe/H]$>$=--1.58. This
absolute magnitude can be compared with $M_V({\rm
RR})$=0.59$\pm$0.37 at [Fe/H]=--1.6 given by eq.[5].  The validity of the
present statistical analysis has been further checked by inspecting
the distribution $(\pi_{\rm H}-\pi_{\rm p})/\sigma_\pi$ for the 125 RR
Lyrae stars.  The median of the distribution is shifted only by 0.08
from the Gaussian.

\section{CONCLUSION}

The present derivation of the absolute magnitude of RR Lyrae stars is
twofold, relying upon Hipparcos proper motions and trigonometric
parallaxes.  First, applying the statistical parallax method to
Hipparcos proper motions of 99 halo RR Lyraes, we have found

\begin{equation}
<M_V({\rm RR})>=0.69\pm0.10 \ \ {\rm at <[Fe/H]>}=-1.58 \ \ \ .
\end{equation}

Second, applying the Lutz-Kelker correction to the most accurate
parallax of HIP95497 (RR Lyr) with $\sigma_\pi/\pi_{\rm H}$=0.135 and
[Fe/H]=--1.37, we have the magnitude range $M_V({\rm
RR})$=(0.62--0.72)$^{+0.28}_{-0.31}$, which yields

\begin{equation}
M_V({\rm RR})=(0.58 - 0.68)^{+0.28}_{-0.31} \ \ {\rm at \ \
[Fe/H]}=-1.6 \ \ \ ,
\end{equation}

\noindent given the slope 0.20 of the $M_V({\rm RR})$-[Fe/H] relation.
The lower and upper bounds in eq.[7] correspond to the assumed spatial
density of RR Lyraes $n(r)$=const.~and $n(r)\propto r^{-2}$,
respectively. Furthermore, applyng the maximum likelihood algorithm to
Hipparcos parallaxes including low accuracy and negative ones as well
for 125 RR Lyraes with --2.49$\leq$[Fe/H]$\leq$0.07, we have found the
relation (5) which gives
\begin{equation}
M_V({\rm RR})=0.59\pm0.37 \ \ {\rm at \ \ [Fe/H]}=-1.6 \ \ \ .
\end{equation}

\noindent Applying the same method to the 99 halo RR Lyraes, we have found 

\begin{equation}
<M_V({\rm RR})>=0.65\pm0.33 \ \ {\rm at \ \ <[Fe/H]>}=-1.58 \ \ \ .
\end{equation}

\noindent Although the trigonometric parallax results have large
errors, these results are consistent with the statistical parallax
result. All of the four results suggest the fainter absolute
magnitude, $M_V({\rm RR})$$\approx$0.6--0.7 at [Fe/H]=--1.6.  Note that three
estimates from parallaxes eqs.[7]-[9] are not completely independent,
in regorous sense.

Recently, Feast \& Catchpole (1997) find $M_V({\rm RR})$ = 0.25 at
[Fe/H]=--1.9, combining the LMC distance modulus of 18.7 determined by
Hipparcos parallaxes of Cepheids with the data of LMC RR Lyraes by
Walker (1992). On the other hand, Ried (1997) estimates the distance
scale to the globular clusters by subdwarf main-sequence fitting on
the bais of Hipparcos parallaxes, giving $M_V({\rm RR})$=0.15 at
[Fe/H]=--2.1. These bright values of $M_V({\rm RR})$ imply much
younger age for the oldest globular clusters than the previous
estimates (Bolte \& Hogan 1995; Chaboyer et al.~1996). However, the
direct estimates of $M_V({\rm RR})$ in the present study support much
fainter value than the indirect estimates obtained by them. The
present results give $M_V({\rm RR}) \approx 0.54-0.64$ at
[Fe/H]=--1.9, given the slope 0.20 of the $M_V({\rm RR})$-[Fe/H]
relation. Note that the Walker's $M_V({\rm RR})$ combined with the LMC
distance modulus of 18.37 (Gould 1995) derived from the SN 1987A ring
distance yields $M_V({\rm RR})$=0.57 at [Fe/H]=--1.9, which favors the
present results. Our findings imply that the age of the oldest
globular clusters in our Galaxy still conflicts with the standard
cosmological model of a flat, matter-dominated universe with the
Hubble constant $H_0=60-80$km$\,$s$^{-1}\,$Mpc$^{-1}$ estimated from
almost all observations (Bolt \& Hogan 1995).  It appears that the age
problem remains still unresolved.

\acknowledgements

This work has been supported in part by the grant-in-Aid for
Scientific Research (08640336) and Center of Excellence (COE) research
(07CE2002) of the Ministry of Education, Science, and Culture in
Japan.  We would like to thank H. Saio for many fruitful discussions.

\end{document}